\documentclass[epjCONF]{svjour}
\usepackage{graphics}
\usepackage[varg]{txfonts} 
\usepackage[latin1]{inputenc}
\session-title{The Space Photometry Revolution - CoRoT Symposium 3, Kepler KASC-7 joint meeting}
\begin{document}
\title{Seventy new non-eclipsing BEER binaries discovered in CoRoT lightcurves and confirmed by RVs from AAOmega}
\author{Lev Tal-Or \inst{1}\fnmsep\thanks{\email{levtalo@post.tau.ac.il}} 
\and Simchon Faigler\inst{1}
\and Tsevi Mazeh \inst{1}}
\institute{School of Physics and Astronomy, Raymond and Beverly Sackler Faculty of Exact Sciences, Tel Aviv University, Israel}
\abstract{We applied the BEER algorithm to the CoRoT lightcurves from the first five LRc fields and identified $481$ non-eclipsing BEER candidates with periodic lightcurve modulations and amplitudes of $0.5-87$\,mmag. Medium-resolution spectra of $281$ candidates were obtained in a seven-night AAOmega radial-velocity (RV) campaign, with a precision of $\sim1$\,km/s. The RVs confirmed the binarity of $70$ of the BEER candidates, with periods of $0.3-10$\,days.
} 

\maketitle
\section{Introduction}
\label{intro}
BEER \cite{Faigler2011} is a new method to discover short-period \textit{non-eclipsing} binaries by using precise photometric lightcurves obtained with space missions like CoRoT and \textit{Kepler}. The BEER algorithm searches for stars that show in their lightcurves a combination of the BEaming, Ellipsoidal, and Reflection modulations induced by a short-period companion.

Recently we have reported several discoveries made by applying the BEER algorithm to \textit{Kepler} data -- seven new binaries with low-mass companions \cite{Faigler2012}, and a hot Jupiter with evidence of superrotation \cite{Faigler2013,Faigler2015}. We report here the discovery and confirmation of $70$ non-eclipsing BEER binaries in CoRoT data.

\section{Observations and data analysis}
\label{sec1}
We applied the BEER algorithm to the white lightcurves of the CoRoT fields LRc01, LRc02, LRc03, LRc04, and LRc05, similarly to the way it was done for \textit{Kepler} lightcurves by \cite{Faigler2012}, and selected a total of $481$ candidates for RV follow-up from all five fields. $281$ of the selected candidates were eventually observed by the AAOmega multi-object spectrograph on the Anglo-Australian Telescope (AAT).

The observations took place on seven consecutive nights starting on August 02, 2012, and for most of the observed candidates we got six to seven medium-resolution ($R\sim10,000$) AAOmega spectra. Observations and data reduction were performed similarly to previously reported works \cite{Sebastian2012}. Table \ref{tab1} lists the coordinates, magnitudes, photometric ephemeris, and amplitudes of the three BEER effects, for the observed candidates.

\begin{table}
\caption{Coordinates and photometric parameters of the BEER candidates observed at AAOmega$^*$.}
\label{tab1}
\begin{tabular}{lcccrrrrrc}
\hline
\hline
CoRoT ID & RA & Dec & V & Orbital & Orbital & Ellipsoidal & Beaming & Reflection & Conf.\\
 & (deg) & (deg) & (mag) & period & phase & amplitude & amplitude & amplitude & flag\\
 & & & & (day) & (HJD-2451545) & (ppm) & (ppm) & (ppm) & \\
\hline
$105659320$ & $280.4275$ & $ 5.8974$ & $14.8$ & $0.70597$ & $3171.931$ & $-2203$ & $367$ & $1497$ & $1$ \\
 & & & & $0.00032$ & $0.035$ & $11$ & $66$ & $111$ & \\
$105962436$ & $280.9442$ & $ 5.8186$ & $14.2$ & $1.80188$ & $3171.579$ & $-3563$ & $220$ & $-677$ & $1$ \\
 & & & & $0.00062$ & $0.025$ & $31$ & $26$ & $43$ & \\
\hline
\end{tabular}
\\$^*$\,Each line of parameters is followed by a line of uncertainties. Two of the table entries are shown here for guidance regarding its form and content. The table is available in its entirety at ftp://wise-ftp.tau.ac.il/pub/corotAAO.
\end{table}

To derive RVs and errors from the observed spectra we calculated cross-correlation functions (CCFs) of the observed spectra with a set of synthetic Phoenix spectra \cite{phoenix99}. To maximize the CCF peak values we selected for each star a template with an optimum set of spectral parameters (i.e. $T_{\rm eff}$, log\,$g$, $[\rm{m/H}]$, and $v\sin i$). For double-lined spectra we used TODCOR \cite{zm94}, optimizing the primary and secondary templates and the flux ratio between them ($\alpha$). The derived RVs, their uncertainties, and template parameters used to derive these RVs can be found at ftp://wise-ftp.tau.ac.il/pub/corotAAO.

\begin{figure}
\centering
\resizebox{9cm}{!}
{\includegraphics{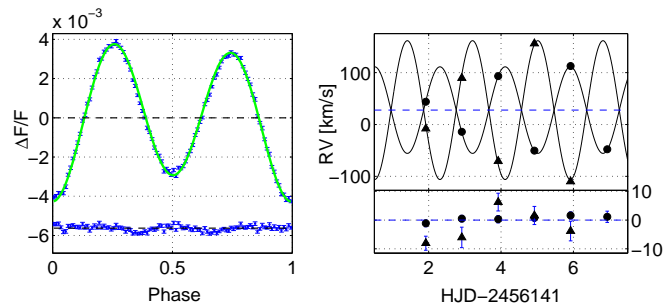}}
\caption{Illustration of the discovery and confirmation process of the BEER SB2 $105962436$. Left: phase-folded lightcurve with the best-fit BEER model. Right: AAOmega RVs with the best-fit Keplerian model.}
\label{fig1}
\end{figure}

To separate between true BEER binaries and false alarms (FAs) we fitted the derived RVs with a circular Keplerian model, taking the BEER period and phase as priors, and applied $\chi^2$- and \textit{F}-statistics for the orbital solution.
For double-line spectra we fitted a circular Keplerian model for the primary and secondary RVs separately, as well as fitting a circular Keplerian model of a double-line binary (SB2) for the two sets of RVs together, in both cases taking the BEER period and phase as priors. Figure \ref{fig1} shows the phase-folded lightcurve with the best-fit BEER model, and the AAOmega RVs with the best-fit Keplerian model for one of the confirmed SB2s -- CoRoT id $105962436$. 

In total, we confirmed the binarity of $70$ non-eclipsing BEER candidates, with $18$ of them being SB2s. The confirmed binaries are indicated in the rightmost column of Table \ref{tab1}.

\section{Conclusions}
\label{conc}
Using RV follow-up observations with AAOmega, we have demonstrated for the first time the capability of the BEER algorithm to detect \textit{non-eclipsing} short-period binaries in CoRoT lightcurves. We discovered $70$ such binaries, with $18$ of them being SB2s. These detections can help revealing the detection rate and the false-alarm statistics of the BEER algorithm applied to CoRoT data.

\end{document}